\documentclass{iau}
\usepackage{graphicx}

\title[S294.~~Helicity transport from convection zone to interplanetary space] 
{Helicity transport from solar convection zone to interplanetary space}

\author[M. Zhang]   
{Mei Zhang}

\affiliation{Key Laboratory of Solar Activity, National Astronomical Observatory, Chinese Academy of Sciences, Datun Road A20, Chaoyang District, Beijing 100012, China \\ email: {\tt zhangmei@bao.ac.cn} \\[\affilskip]}

\pubyear{2013}
\volume{294}  
\pagerange{119--126}
\jname{Solar and Astrophysical Dynamos and Magnetic Activity}
\editors{A.G. Kosovichev, E.M. de Gouveia Dal Pino \& Y. Yan, eds.}
\begin{document}

\maketitle

\begin{abstract}
Magnetic helicity is a physical quantity that describes field topology. It is also a conserved quantity as Berger in 1984 demonstrated that the total magnetic helicity is still conserved in the corona even when there is a fast magnetic reconnection. It is generally believed that solar magnetic fields, together with their helicity, are created in the convection zone by various dynamo processes. These fields and helicity are transported into the corona through solar photosphere and finally released into the interplanetary space via various processes such as coronal mass ejections (CMEs) and solar winds. Here I will give a brief review on our recent works, first on helicity observations on the photosphere and how to understand these observations via dynamo models. Mostly, I will talk about what are the possible consequences of magnetic helicity accumulation in the corona, namely, the formation of magnetic flux ropes, CMEs taking place as an unavoidable product of coronal evolution, and flux emergences as a trigger of CMEs. Finally, I will address on in what a form magnetic field in the interplanetary space would accommodate a large amount of magnetic helicity that solar dynamo processes have been continuously producing.
\keywords{Sun: magnetic fields, Sun: interior, Sun: photosphere, sunspots, Sun: activity, Sun: corona, Sun: coronal mass ejections (CMEs), interplanetary medium}
\end{abstract}

\firstsection 

\section{Introduction}

Magnetic helicity is defined as $H = \int_V {\bf A} \cdot {\bf B} ~dV$, where {\bf B} is the vector magnetic field and {\bf A} is the vector potential of ${\bf B}$ defined by ${\bf B} = \nabla \times {\bf A}$. As a physical quantity magnetic helicity describes field topology, quantifying the twist (self-helicity) and linkage (mutual-helicity) of magnetic field lines.

Figure 1, taken from Wiegelmann \& Sakurai (2012), gives a simple illustration. If the field lines in the torus have a configuration as in the left panel, the field has no magnetic helicity, that is, $H = 0$. If the field lines in the torus have a configuration as in the middle panel, the field has a self-helicity with $H = T \Phi^2$, where $\Phi$ is the total magnetic flux and $T$ is the number of turns the field line twists. If the two untwisted tori link with each other as in the right panel, then the field has a mutual helicity with $H = \pm 2 \Phi_1 \Phi_2$.

\begin{figure}
\begin{center}
 \includegraphics[width=4.8in]{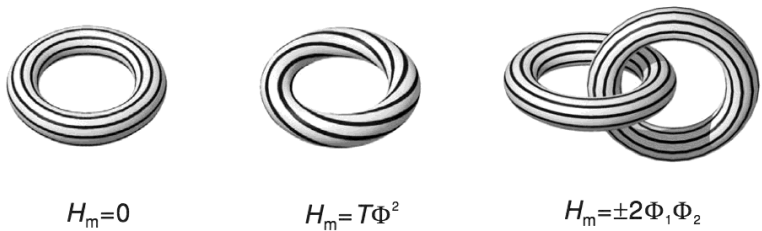}
 \caption{A simple example to illustrate the concepts of magnetic helicity, self helicity (middle panel) and mutual helicity (right panel); taken from Wiegelmann \& Sakurai (2012).}
\end{center}
\end{figure}

What really makes the helicity study important is the property that the total magnetic helicity is still conserved in the corona even when there is a fast magnetic reconnection (Berger 1984). This indicates that once the magnetic helicity is created in the convection zone by dynamo processes and transported into the corona through photosphere, it then cannot be annihilated even by dramatic events such as solar flares. This conservation rule, together with the observed hemispheric helicity sign rule (which we are going to address in the following Sections 2 and 3), naturally leads us to conclude that the total magnetic helicity is being accumulated in the corona. The accumulation of the magnetic helicity in the corona will bring in several important consequences, which we are going to discuss in Section 4.

Furthermore, this conservation rule of total magnetic helicity also tells us that, compared to magnetic energy, magnetic helicity is much less dissipative. It is rather transported and/or redistributed than being dissipated. And with the transport and/or redistribution of magnetic helicity, the magnetic field along with its flux and energy are then also being transported and/or redistributed.

Another interesting point to put on here is that, making use of the total magnetic helicity conservation rule gives us a way to avoid treating the difficult process of magnetic reconnection. Even though we may not know the detailed processes of magnetic reconnection, we know that the total magnetic helicity is approximately conserved before and after flares. This gives us a chance to glimpse on what could be the possible end states of magnetic reconnections. Section 4.1 gives an example of such a type of study.


\section{Hemispheric helicity sign rule observed on the photosphere}

Over the past two decades many observations have shown that magnetic fields, created by dynamo processes in the solar interior, are emerging on the solar photosphere with a preferred helicity sign in each hemisphere, namely, positive helicity sign in the southern hemisphere and negative helicity sign in the northern hemisphere (Pevtsov et al. 1995, 2001; Bao \& Zhang 1998; Hagino \& Sakurai 2004, 2005; Zhang 2006; Wang \& Zhang 2010; Hao \& Zhang 2011). This is the so-called usual hemispheric helicity sign rule. It has been studied and confirmed using data sets obtained with different instruments located in different places of the world including space. It has also been tested using data for four solar cycles (that is, solar cycles 21, 22, 23 and 24).

Note that this hemispheric helicity sign rule is also evident in the global Sun, not just in active regions. Using a technique used by Pevtsov \& Latushko (2000) to reconstruct the vector magnetic field from longitudinal magnetic field observations, Wang \& Zhang (2010) examined the hemispheric helicity sign rule for large-scale magnetic fields. They found that the same hemispheric helicity sign rule presents everywhere in the global magnetic field, extending to 60 degrees high in solar latitudes, and is preserved through the whole solar cycle 23.

\begin{figure}
\centerline{\includegraphics[width=12cm]{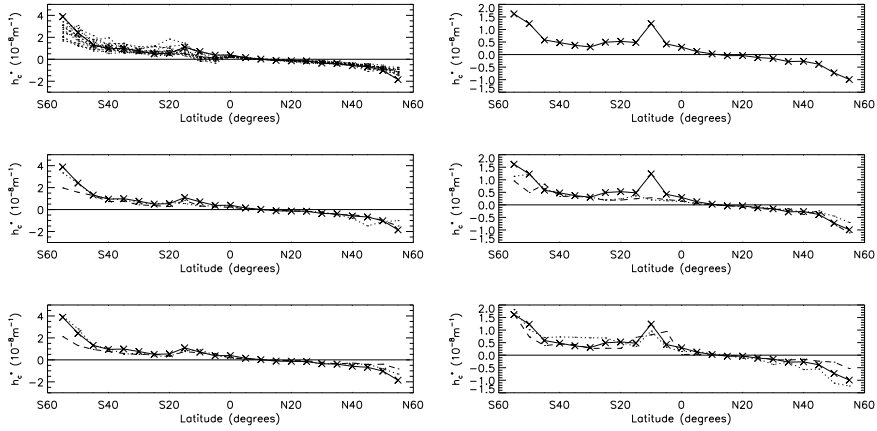}}
\caption{\small{Variations of normalized longitudinally-averaged current helicity with the latitude. Profiles in left panels are obtained by using MDI/SOHO data and the right panels KPVT/NSO data. See text and Wang \& Zhang (2010) for details.}}
\end{figure}

Figure 2 shows the normalized profiles of current helicity in a Carrington rotation starting at September 14, 1996. Left panels shows the profiles obtained by using MDI/SOHO data and the right panels KPVT/NSO data. Different solid and dotted lines in each panel represent the profiles obtained by using different magnetograms and different reconstruction parameters, details of which are described in Wang \& Zhang (2010). The point to show here is that all these profiles show a clear and consistent pattern that follows the established hemispheric helicity sign rule, independent of the instruments and the parameters used.

Complication actually comes in with active regions. Using so-far the most accurate vector magnetic field measurements obtained with SP/Hinode, Hao \& Zhang (2011) found that the usual hemispheric helicity sign rule is preserved in the ascending phase of solar cycle 24, but not in the descending phase of solar cycle 23. They also found that, in two example sunspots, the helicity parameters change their signs from the inner umbra to the outer penumbra. This is consistent with Zhang (2006)'s finding that strong and weak fields in active regions possess opposite helicity signs, a statistical result from the analyze of a large sample of 17,200 vector magnetograms.


\section{Helicity production in a convective Babcock-Leighton dynamo model}

Figure 3 presents a very preliminary result of our recent work on using a convective Babcock-Leighton dynamo model (Miesch \& Brown 2012) to understand the helicity production processes in solar dynamos.

Using the anelastic spherical harmonic (ASH) code and taking into account the influence of magnetic flux emergence by means of the Babcock-Leighton mechanism, the three-dimensional simulation of solar/stellar convection in Miesch \& Brown (2012) has successfully produced the cyclic activity reminiscent of solar observations such as the equatorward propagation of toroidal flux. This is illustrated in the left panel of Figure 3, where toroidal magnetic field is shown as the black (negative) and white (positive) in the image, the X-axis is the time and the Y-axis the solar latitude.

Similar to the left panel, the magnetic helicity density (i.e. ${\bf A} \cdot {\bf B}$) and current helicity density (i.e. $ (\nabla \times {\bf B}) \cdot {\bf B}$) are shown in the middle and right panels respectively. We can see from these maps that both magnetic helicity density and current helicity density show cyclic variation in their magnitudes. Hemispheric helicity sign rule shows up clearly in the magnetic helicity density map, consistent with the observed usual hemispheric helicity sign rule. Whereas the current helicity density map also shows a pattern consistent with the usual hemispheric helicity sign rule, it is less pronounced than that in the magnetic helicity density map. Also interesting to notice is that current helicity density map does show opposite-sign patches at one time in one hemisphere, consistent with our observations mentioned above.

\begin{figure}
\centerline{\includegraphics[width=15cm]{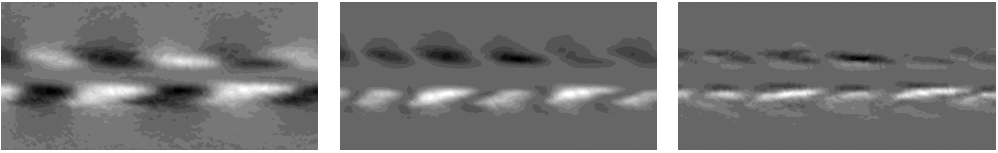}}
\caption{\small{Cyclic variations of toroidal magnetic field (left panel), magnetic helicity density (middle panel) and current helicity density (right panel) in a convective Babcock-Leighton dynamo model.}}
\end{figure}


\section{Consequences of magnetic helicity accumulation in the corona}

 As we have mentioned above, the conservation rule of total magnetic helicity together with the observed hemispheric helicity sign rule naturally lead to the accumulation of total magnetic helicity in the corona. In this section, we discuss from a theoretical point of view on what are the possible consequences of magnetic helicity accumulation in the corona.

\subsection{Formation of magnetic flux ropes}

The first consequence of magnetic helicity accumulation in the corona is the formation of magnetic flux ropes.

The solar corona is a body of such tenuous plasma that we usually regard its large-scale magnetic field as force-free. Woltjer Theorem (Woltjer 1958) states that, if the total magnetic helicity conservation is the only constraint, then of all the magnetic fields sharing the same amount of total magnetic helicity, one of the force-free fields with a constant-alpha is the absolute minimum-energy state. This minimum-energy state can be reached through turbulent magnetic reconnections, a process termed as Taylor relaxation (Taylor 1974).

Figure 4 presents the result of a calculation of a simple model to illustrate the formation of a flux rope through this process of Taylor relaxation (Zhang \& Low 2003). The calculation relates two axisymmetric force-free fields as a higher-energy initial state (left panel) and a lower-energy end-state (right panel), where during the transition between the two states the boundary flux distribution is unchanged and the total magnetic helicity is conserved. Comparing the two states we see that a toroidal magnetic flux rope (illustrated in the figure as the structure detached from the inner boundary of the domain) has formed from the initial field that contains no magnetic flux rope.

The key physics here is that, when a certain amount of magnetic helicity has been transported into the corona, Taylor relaxation involves a redistribution of the conserved magnetic helicity and puts a main part of the accumulated helicity into the flux rope. In other words, once a Taylor relaxation is allowed, forming a magnetic flux rope becomes the most efficient way for the field to store magnetic helicity.

\begin{figure}
\centering
\begin{minipage}[t]{0.3\textwidth}
{\includegraphics[width=0.85\textwidth]{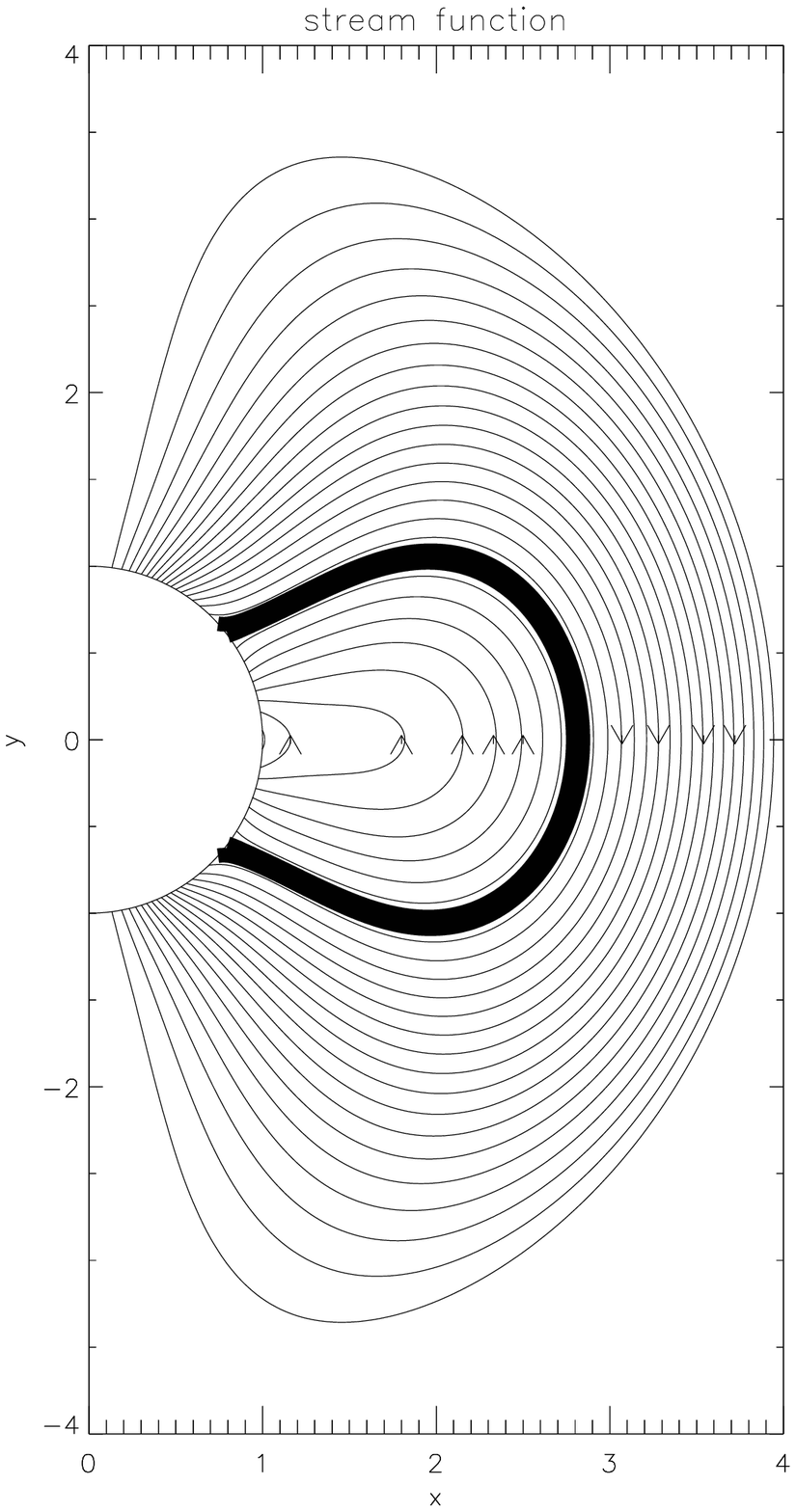}}
\end{minipage}
\begin{minipage}[t]{0.3\textwidth}
{\includegraphics[width=0.85\textwidth]{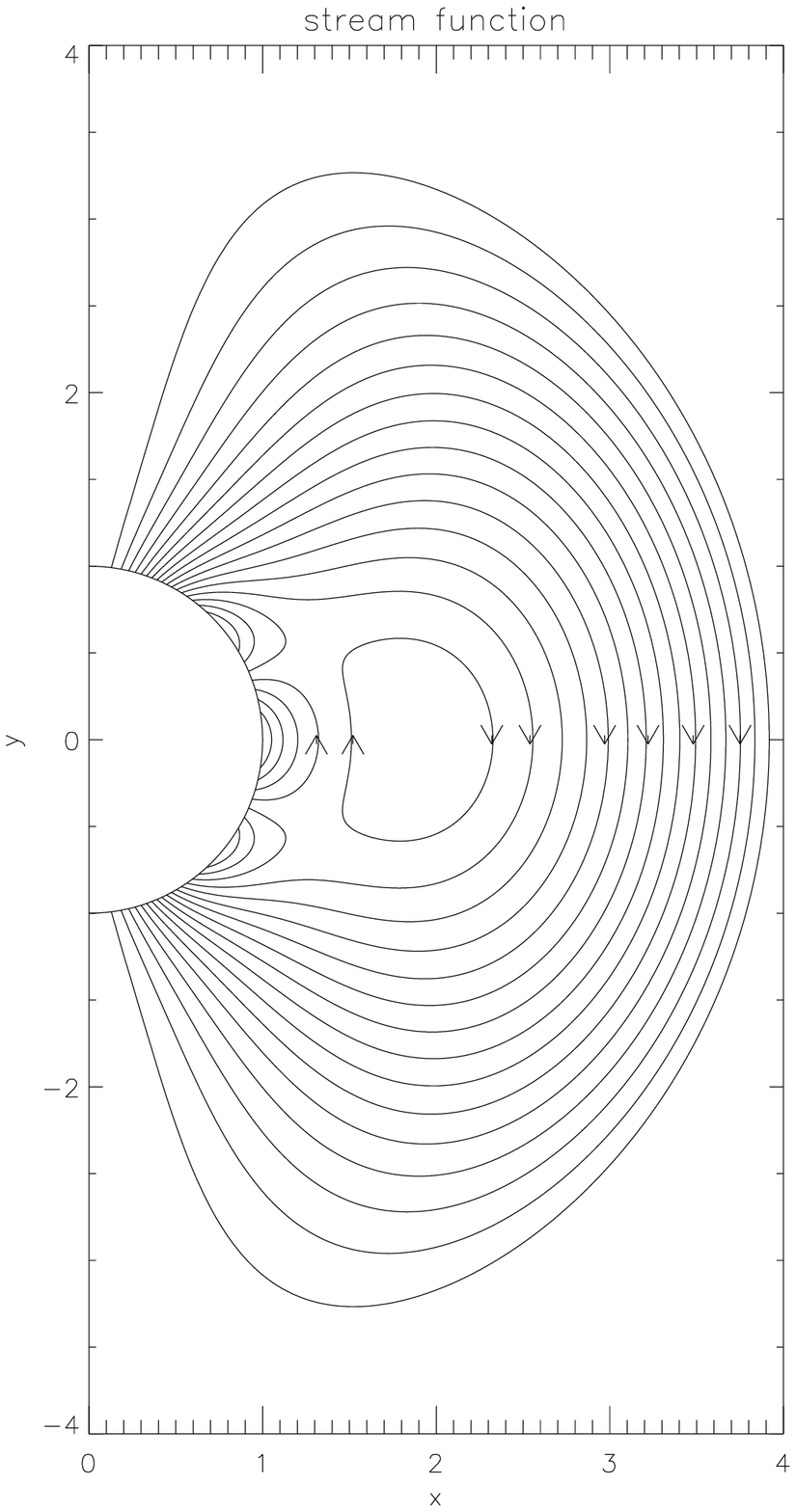}}
\end{minipage}
\caption{\footnotesize{Contours of stream functions of an initial, current-sheet force-free field (left panel) and the end-state, smooth force-free field (right panel), taken from Zhang \& Low (2003). The right panel indicates that a magnetic flux rope has formed in the end state as a result of Taylor relaxation.}}
\end{figure}

\subsection{CMEs taking place as an unavoidable product of coronal evolution}

The second consequence of magnetic helicity accumulation in the corona is that solar eruptions such as CMEs become unavoidable.

With magnetic helicity being continuously transported into the corona through solar photosphere, the coronal magnetic field contains more and more magnetic helicity. However, as Zhang et al. (2006) first pointed out, for a given boundary flux distribution, there is an upper bound on the total magnetic helicity that force-free fields can contain. Once the accumulated magnetic helicity exceeds this upper bound, an eruption becomes unavoidable.

Figure 5 gives an example of their calculations. Each plus symbol in the figure represents a force-free field, all having the same boundary flux distribution as that of a dipolar field. Using some specific mathematical methods they were able to get a complete set of axisymmetric power-law force-free fields for each specific $n$ index (see Flyer et al. 2004 for mathematical treatments and Zhang et al. 2006 for more physical discussions). The solutions of these force-free fields suggest that there may be an upper bound on the total magnetic helicity that force-free fields can contain.

\begin{figure}
\begin{center}
 \includegraphics[width=3.in]{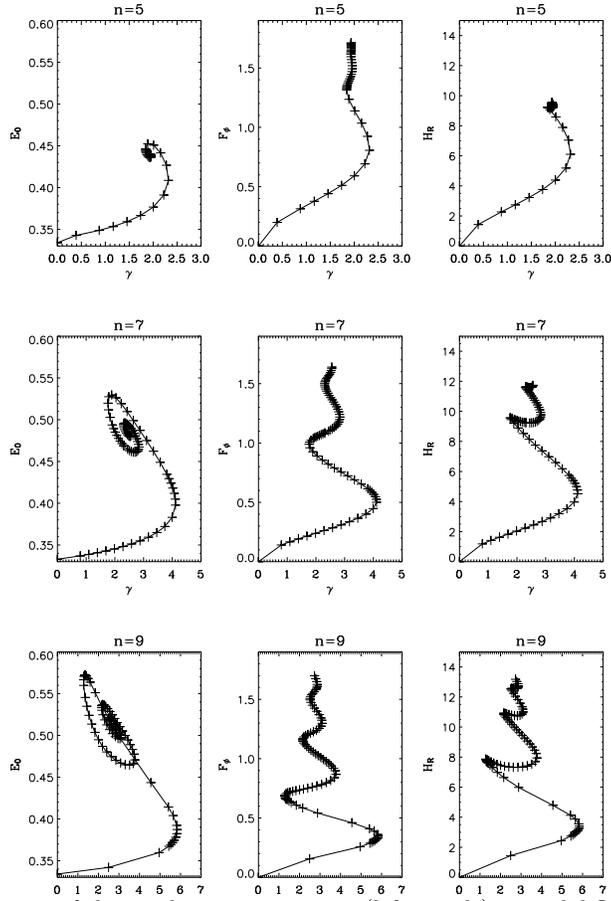}
 \caption{Variation of the total magnetic energy (left panels), toroidal flux (middle panels) and total magnetic helicity (right panels) along the solution curve, taken from Zhang et al. (2006). These solutions of force-free fields suggest that there are upper bounds on all three quantities, i.e. total magnetic energy, total toroidal flux and total magnetic helicity.}
\end{center}
\end{figure}

The essence of the existence of a helicity upper bound is that a certain amount of poloidal magnetic flux can only contain a certain amount of toroidal flux. Roughly speaking, magnetic helicity is a product of poloidal flux and toroidal flux. As shown in Figure 6, a potential field with purely poloidal flux (left panel) is stable and a field with purely toroidal flux cannot exist (middle panel). General force-free fields can be viewed as a combination of these two (right panel), that is, a toroidal field is confined by surrounding poloidal field. If we increase the toroidal field continuously, which is equivalent to increasing the magnetic helicity continuously, then up to a point, comparing to the torodial field the surrounding poloidal field becomes so weak that the field becomes similar to that in the middle panel. At this moment no further equilibrium exists and an eruption becomes unavoidable.

If this picture is true, then this means that with the continuous accumulation of magnetic helicity in the corona, an expulsion such as a CME will become unavoidable, as a natural result of coronal evolution.

\begin{figure}
\begin{center}
 \includegraphics[width=5.0in]{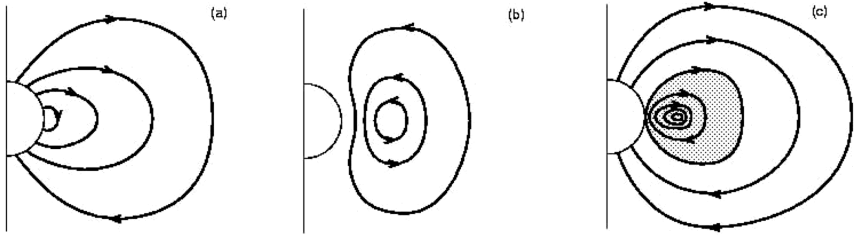}
 \caption{A carton illustration on the essence of the existence of a helicity upper bound, revised from Figure 1 in Zhang et al. (2006). See text for explanations.}
\end{center}
\end{figure}

\subsection{Flux emergences as a trigger of CMEs}

Solving the same governing equation as in Zhang et al. (2006), Zhang \& Flyer (2008) further studied force-free fields with two new boundary conditions. One new boundary condition has a form representing bipolar active regions where the magnetic fluxes are more concentrated near the equator. These fields are referred to as bipolar fields. The other new boundary condition has a form of representing complicated active regions where multipolar fields exist within one active region. These fields are referred to as multipolar fields. For a given boundary condition Zhang \& Flyer (2008) still found the existence of an upper bound on the total magnetic helicity.

However, their calculations also show that the magnitude of the upper bound of total magnetic helicity non-trivially depends on the boundary flux distribution. The upper bound of bipolar fields is lower than that of dipolar fields (those fields in Zhang et al. 2006) and the upper bound of multipolar fields is 10 times smaller than that of dipolar fields.

These results can be used to understand flux emergence triggered or other boundary variation associated CMEs. Imagine the Sun has a corona that has accumulated a certain amount of magnetic helicity but has not yet arrived the criteria for an eruption, that is, the accumulated magnetic helicity has not exceeded the upper bound of force-free fields yet. Then, new magnetic flux emerges on the photosphere. This is equivalent to a change of boundary flux distribution. The new upper bound of total magnetic helicity, corresponding to the new boundary flux distribution, may be lower than that of the previous one, making the already-accumulated total magnetic helicity exceeding the new upper bound. At this moment an eruption becomes ready to take place.

These results can also explain why complicated active regions such as delta sunspots are more ready for eruptions.
Note that complicate active regions have a field topology like multipolar fields. Calculations show that the upper bound of multipolar fields can be 10 times smaller than that of dipolar fields. This means that fields with this type of field topology need much less amount of total magnetic helicity for eruptions. This explains why on the Sun complicated active regions are much easier to erupt because they need less accumulated total magnetic helicity to reach the helicity upper bound.

It is worthy of emphasizing here that the role of changing boundary flux distribution will in no case take away the importance of helicity accumulation. Even though the new helicity upper bound may become lower because of a changed boundary, the already accumulated helicity still needs to exceed the upper bound of the new boundary for an eruption. As an example, Zhang et al. (2008) found that although 91\% of 189 CME-source regions are found to have small-scale flux emergence the same percentage of small-scale flux emergence can also be identified in active regions during periods when there are no solar activities such as flares and CMEs. This means that the flux emergence alone is not a sufficient condition for eruptions.

\subsection{Understanding CMEs in terms of magnetic helicity accumulation}

CMEs are a major form of solar activity. A CME takes away a body of plasma from the low corona into the solar wind and disturbs the near-Earth space if the CME is earth-directed. Since their first systematic detections in the 1970s, thousands of CMEs have been observed and various models have been proposed (e.g. see Low 2001, Zhang \& Low 2005 for a review). Here we provide and summarize our understandings on CMEs in terms of magnetic helicity accumulation, in relation to four questions posed by CME observations.

1) {\it Why do CMEs take place?} This is because the corona has accumulated enough total magnetic helicity for eruptions.

2) {\it Why do CMEs take place occasionally, not continuously?} This is because the helicity transport process, determined by dynamo processes in the convection zone, is slow and the corona needs time to accumulate enough total magnetic helicity for eruptions.

3) {\it Why do CMEs erupt from previously closed regions such as active regions or streamers?} This is because these closed regions are where magnetic helicity can be accumulated. We imagine that magnetic helicity is also being transported into the corona in open regions such as coronal holes. But the transported helicity will immediately propagate along the field lines to interplanetary space and will not get accumulated in the corona.

4) {\it Why do CME initiations often associate with surface field variations such as flux emergences?} This is because with the changed boundary flux distribution the helicity upper bound may be reduced, making the already accumulated total magnetic helicity exceeding the new upper bound.

Our understandings on CMEs provide a vivid model that can be used for space weather prediction. If we can monitor how much magnetic helicity has been accumulated in the corona and at the same time know what is the upper bound on the total magnetic helicity of the current boundary flux distribution, then we can in principle predict whether an eruption will take place or not. However, this requests an accurate measurement of helicity transportation as well as a fast computation of the helicity upper bound for the ever-changing boundary distribution. Many issues, including the accurate measurement of magnetic field on the photosphere (e.g. Wang et al. 2009a, 2009b), need better understandings before we can reach the goal of a reliable prediction.


\section{Magnetic helicity in the interplanetary space}

Zhang et al. (2012) further studied in what a form the magnetic field in the interplanetary space would accommodate a large amount of magnetic helicity that has been brought into the interplanetary space via CMEs and solar winds.

They used the same mathematical treatment as in Zhang et al. (2006) and Zhang \& Flyer (2008), but this time they used the boundary flux distribution defined by Low \& Lou (1990) self-similar force-free solutions. They first found that within the same family of axisymmetric power-law force-free fields that have the same boundary flux distribution and the same power-law index $n$, the self-similar force-free field is always the end-state that possesses the maximum total magnetic helicity. This not only shows that self-similar force-free fields are good magnetic helicity containers but also gives an indication to why self-similar fields are often found in the astrophysical systems where magnetic helicity accumulation is presumably also taking place.

This finding also provides them a physical base to make use of the semi-analytical solutions of self-similar force-free fields in Low \& Lou (1990) to study the evolution of force-free fields, particularly when the field has accumulated a huge amount of magnetic helicity, a time that the numerical treatment has run out of resource. Figure 7 shows the result of such an evolution. We see that as $n$ increases (equivalent to total magnetic helicity of the field increases), more and more field lines become open. When $n$ reaches 201 (i.e. a huge amount of magnetic helicity has been stored in the field), the field lines, projected on the $r-\theta$ plane, become almost everywhere radial, approaching to those in a fully opened-up field such as the potential fully-open Aly field.

\begin{figure}
\begin{center}
 \includegraphics[width=3.2in]{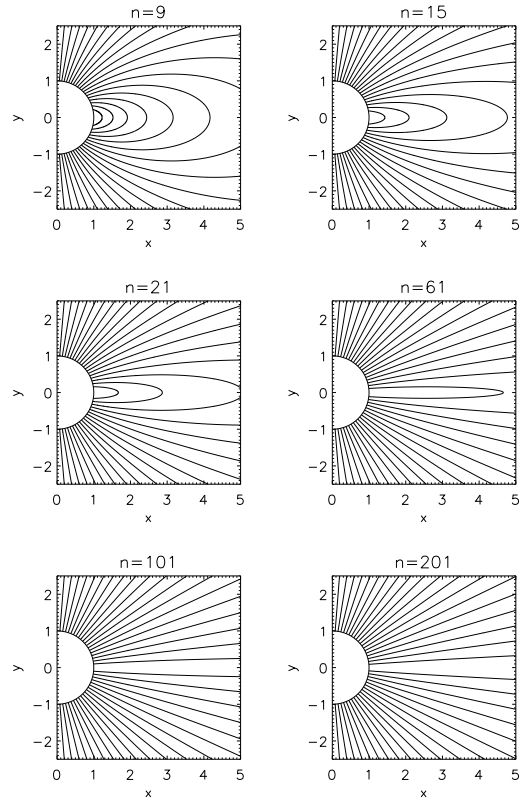}
 \caption{Evolution of field configurations of self-similar force-free fields. We see here that the field lines become more and more open as $n$ (and magnetic helicity) increases.}
\end{center}
\end{figure}

Note that even though the 2D field lines look like radial, the 3D field lines are actually not. Figure 8 shows a comparison of the 3D field lines between the $n=201$ self-similar force-free field (purple lines) and the potential fully-open Aly field (blue lines). From top to bottom panels, plotted respectively are the field lines $0.5^\circ$, $1^\circ$, $2^\circ$ and $20^\circ$ away from the equator. The equator is located in the X-Y plane and the poles in the Z direction. The length of each axis in the left panels is 100 solar radius and the length of each axis in the right panels is 5 solar radius. The central red sphere shows the size of the Sun. We see that, whereas the field lines in the potential Aly field are all radial, the field lines in the $n=201$ self-similar force-free field show impressive Parker-spiral-like structures. This is quite understandable because in one case the field is potential and has no magnetic helicity, whereas in the other case there is a huge amount of total magnetic helicity stored in the field. The field lines between these two fields become undistinguishable only if they originate from latitudes larger than $20^\circ$ from the equator (where the current sheet lies in). Due to these results, we suggest that the existence of Parker-spiral-like structures in the field is the way how the magnetic field in the interplanetary space possesses a huge amount of magnetic helicity and yet, at the same time, keeps all its field lines fully open to infinity.

\begin{figure}
\centering
\begin{minipage}[t]{0.35\textwidth}
{\includegraphics[width=0.75\textwidth]{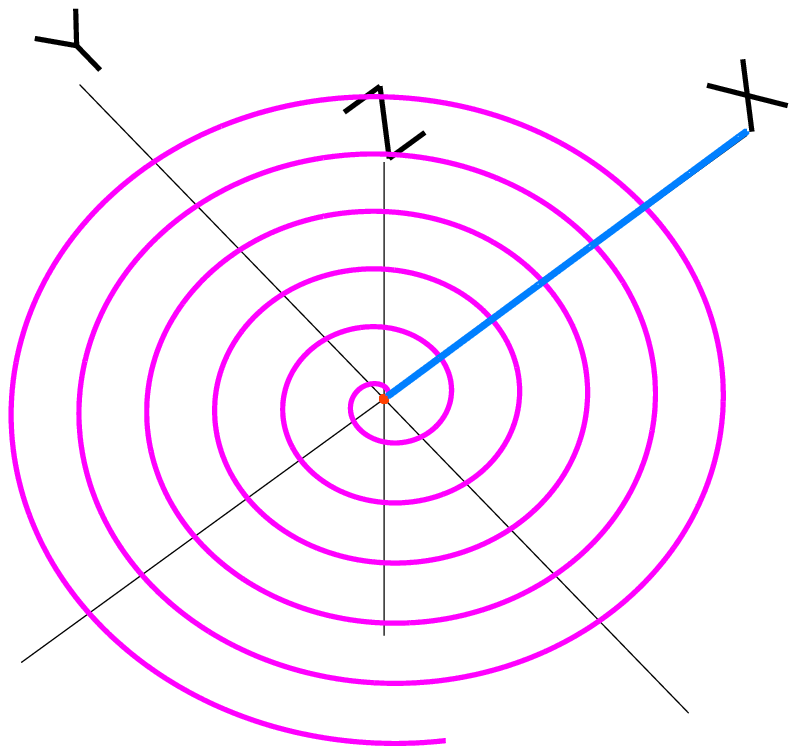}}
\end{minipage}
\begin{minipage}[t]{0.35\textwidth}
{\includegraphics[width=0.75\textwidth]{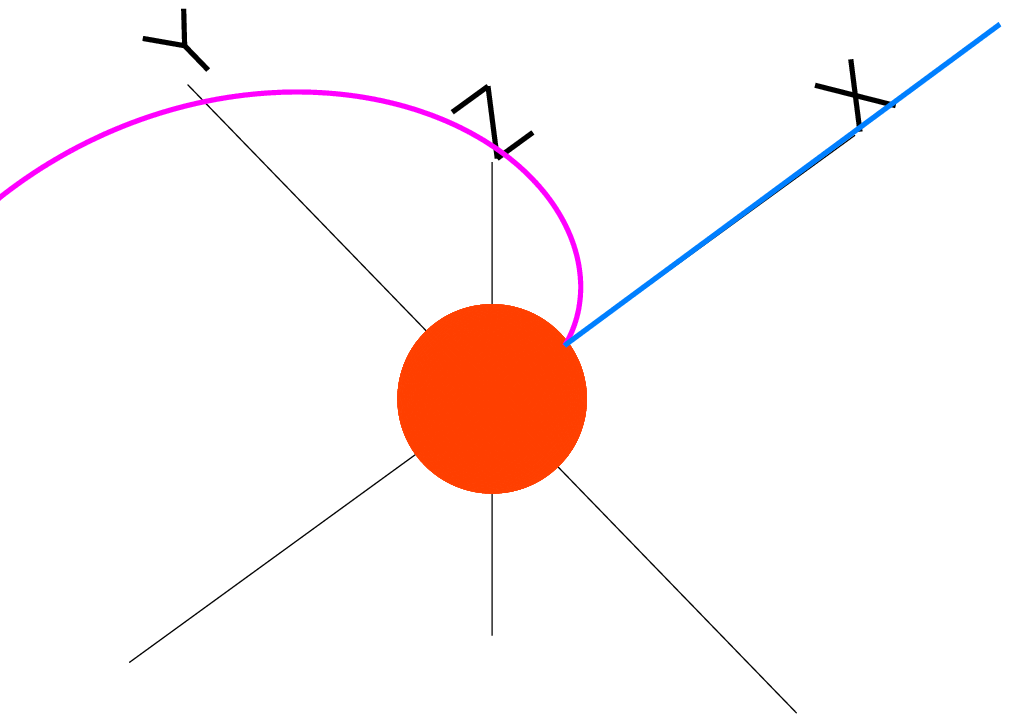}}
\end{minipage}
\begin{minipage}[t]{0.35\textwidth}
{\includegraphics[width=0.75\textwidth]{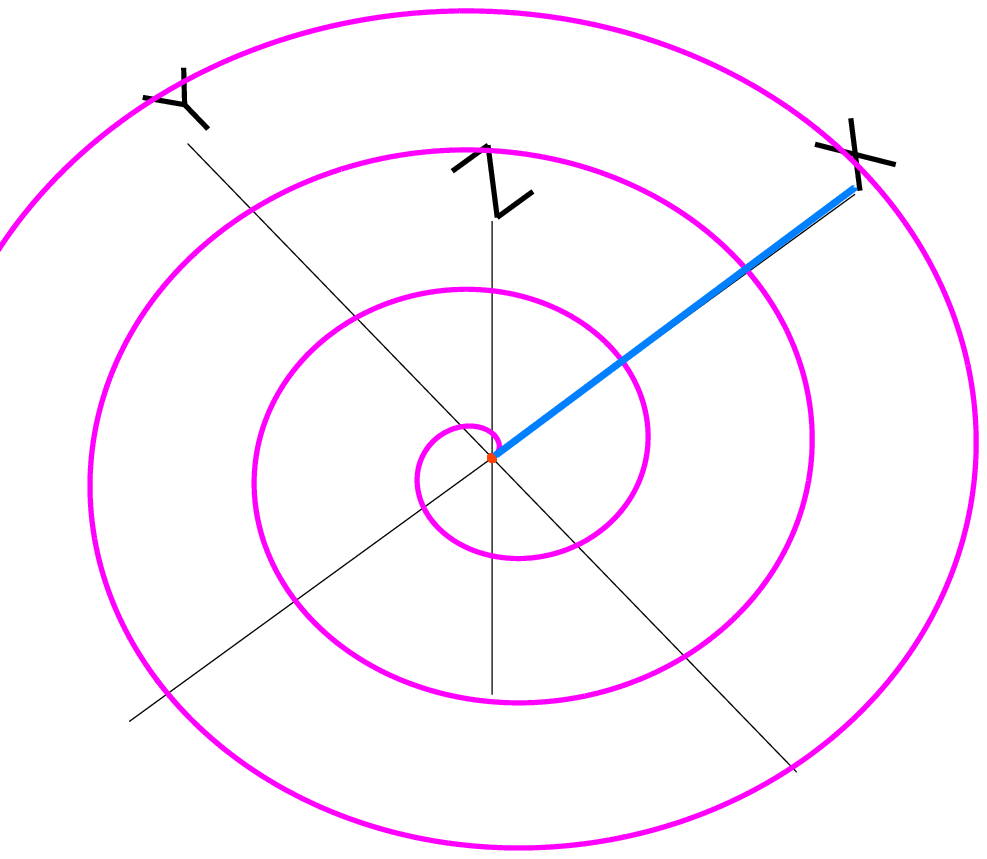}}
\end{minipage}
\begin{minipage}[t]{0.35\textwidth}
{\includegraphics[width=0.75\textwidth]{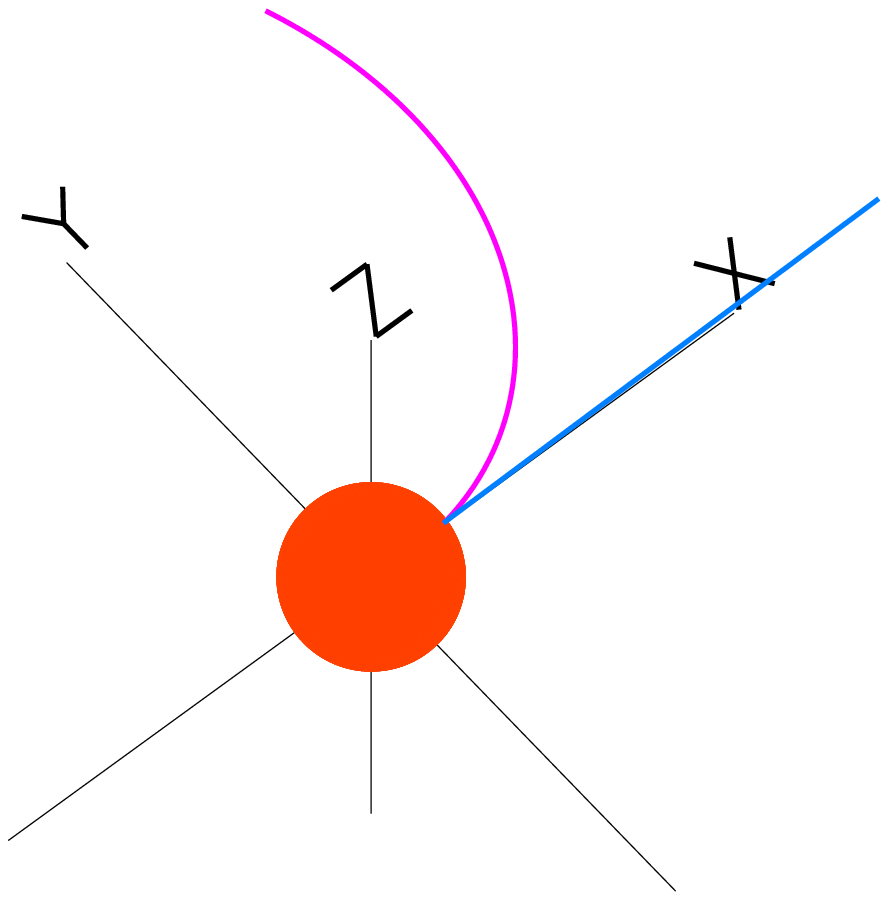}}
\end{minipage}
\begin{minipage}[t]{0.35\textwidth}
{\includegraphics[width=0.75\textwidth]{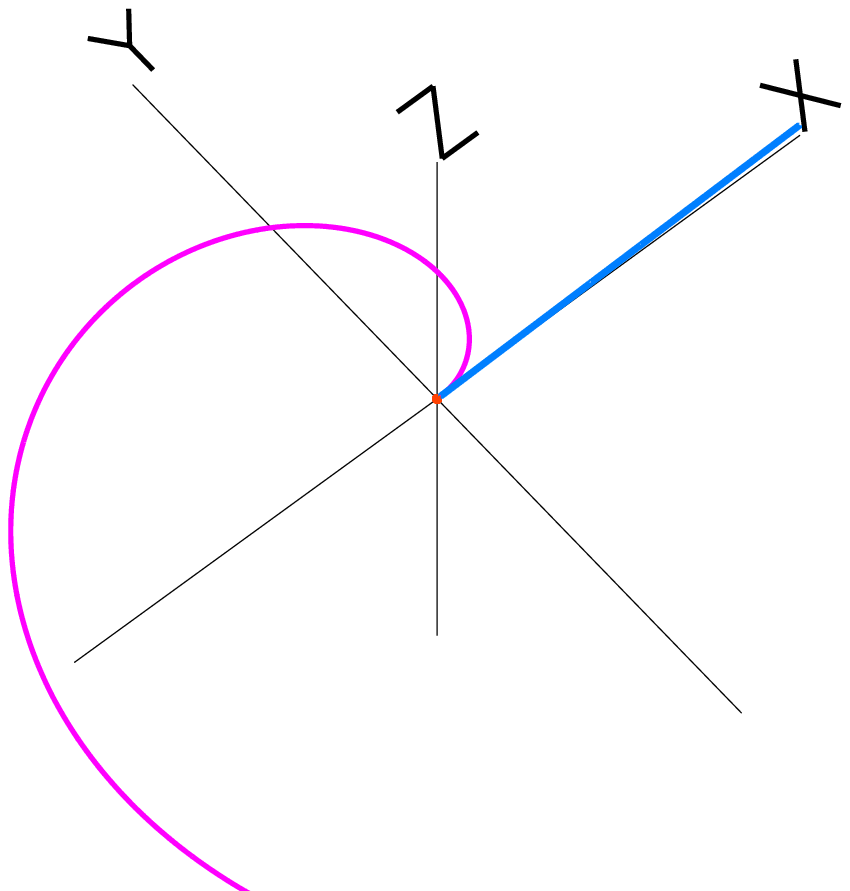}}
\end{minipage}
\begin{minipage}[t]{0.35\textwidth}
{\includegraphics[width=0.75\textwidth]{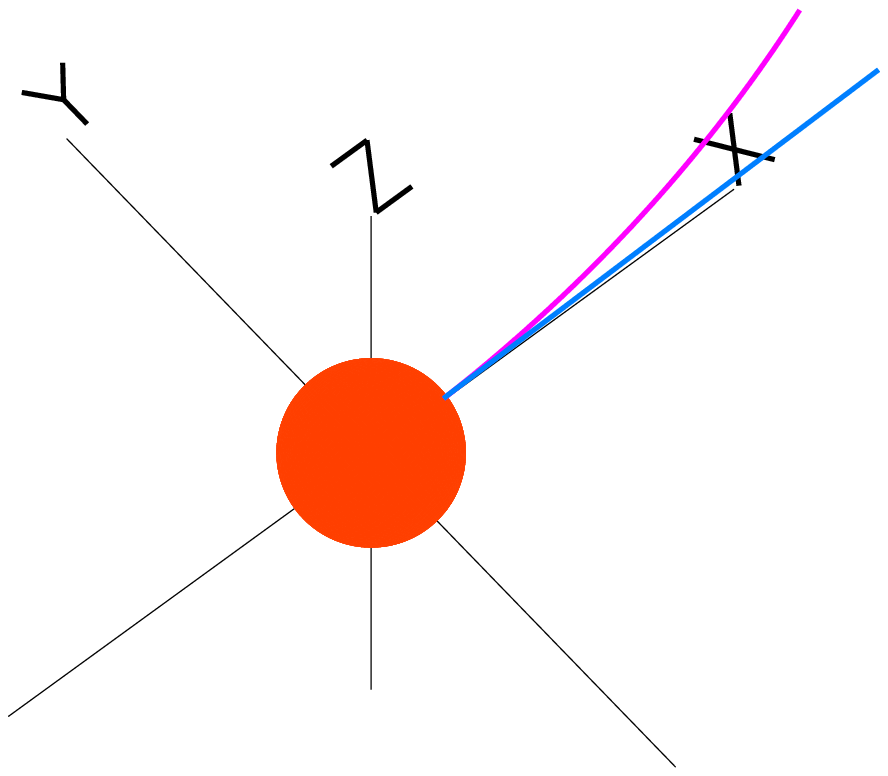}}
\end{minipage}
\begin{minipage}[t]{0.35\textwidth}
{\includegraphics[width=0.75\textwidth]{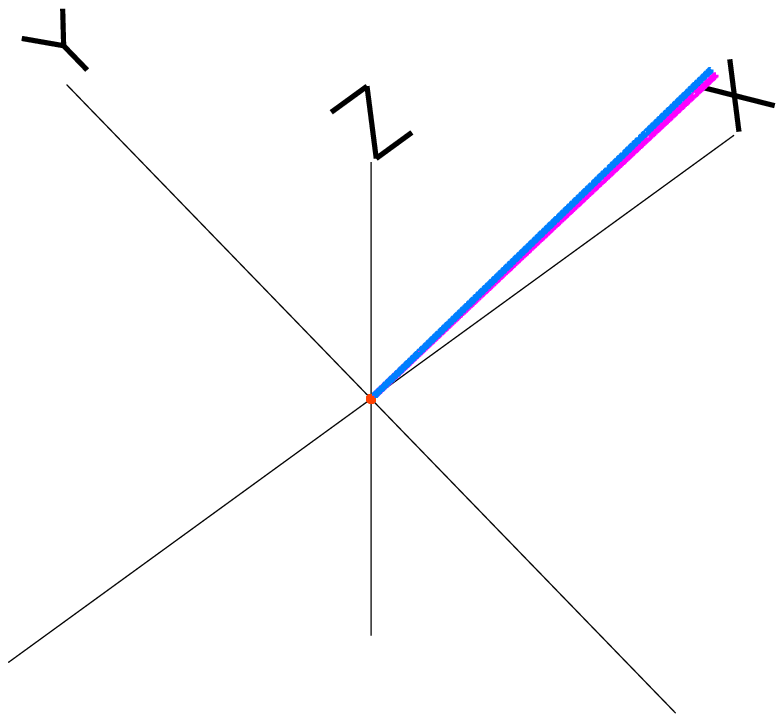}}
\end{minipage}
\begin{minipage}[t]{0.35\textwidth}
{\includegraphics[width=0.75\textwidth]{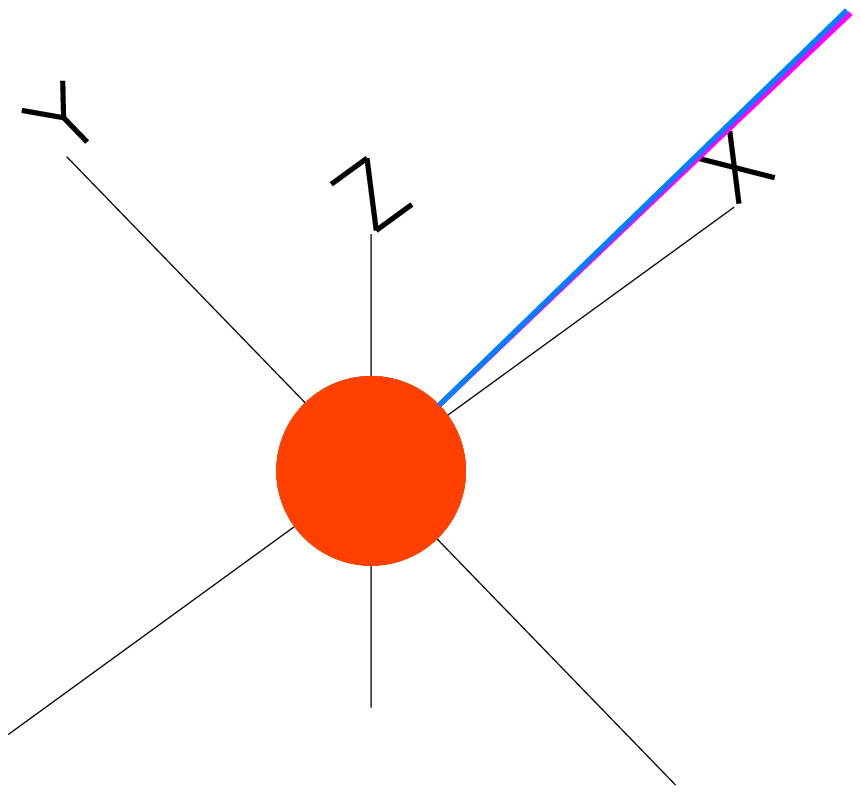}}
\end{minipage}
\begin{minipage}[b]{\textwidth}
\caption{\footnotesize{A comparison of 3D field lines between the $n=201$ self-similar force-free field (purple lines) and the potential fully-open Aly field (blue lines).}}
\end{minipage}
\end{figure}


\section{Concluding remarks}

We conclude with following points.\\
1. Hemispheric helicity sign rule is observed on the photosphere, in both global Sun and in active regions.\\
2. Dynamo models (at least one of them that checked) produce magnetic field consistent with the observed hemispheric helicity sign rule. And it is found that the hemispheric helicity sign rule is better preserved in magnetic helicity density than in current helicity density.\\
3. The accumulation of magnetic helicity in the corona can give rise to magnetic flux rope formation in the corona. It will also result in CMEs as a natural product of coronal evolution.\\
4.	When a huge amount of magnetic helicity is dumped into the interplanetary space, Parker-spiral-like structures will form.

\vspace{5mm}

{\bf Acknowledgements}

We acknowledge supports of the National Natural Science Foundation of China (Grants No. 10921303 and No. 11125314), the National Basic Research Program of MOST (Grant No. 2011CB811401) and the Knowledge Innovation Program of the Chinese Academy of Sciences (Grant No. KJCX2-EW-T07).






\end{document}